\documentclass[showpacs,aps,prd]{revtex4}
\newcommand{\etal}    {{\it et al.}}

\def\Thc     {\ensuremath{\Theta_c^{0}}\xspace}

\usepackage{graphicx}

\newcommand{\BABARPubYear}    {05}
\newcommand{\BABARPubNumber}  {072}

\newcommand{\SLACPubNumber} {11454}

\input babarsym

\begin{document}

\begin{flushright}
\babar-PROC-\BABARPubYear/\BABARPubNumber\\
SLAC-PUB-\SLACPubNumber\\
\end{flushright}

\title{Searches for Pentaquark Baryons at \babar\ \footnote{Work supported by Department of Energy contract DE-AC02-76SF00515.}}


\author{Tetiana Berger-Hryn'ova}
\affiliation{Stanford Linear Accelerator Center, 2575 Sand Hill Road, Menlo Park, CA 94025, USA}
\author{Representing the \babar\ Collaboration}
\noaffiliation

\begin{abstract}
This paper presents the results of inclusive searches for the strange pentaquark states $\Theta^+(1540)$, 
$\Xi_5^{--}(1860)$ and $\Xi_5(1860)^{0}$ as well as the anti-charm pentaquark state $\Theta_c(3099)^{0}$ in a dataset of
123.4 fb$^{-1}$ collected on and 40$\,$MeV below the $\Upsilon(4S)$ resonance by the \babar\ detector at the 
$e^+e^-$ \pep2\ storage rings. No evidence for the pentaquark states is found and upper limits on the rate of 
$\Theta^+(1540)$ and $\Xi_5^{--}(1860)$ production in $e^+e^-$ annihilation are obtained.
\end{abstract}

\maketitle

\begin{center}
{\it Submitted to XIth International Conference on Hadron Spectroscopy, \\
Rio de Janeiro, Brazil, August 21--26 2005}
\end{center}

\section{Introduction} 
\label{intro} 

In the past two years 
several experimental groups have reported observations of a
new, manifestly exotic (B$=$1, S$=$1) baryon resonance, the $\Theta^+$(1540)~\cite{theta}, with
an unusually narrow width ($\Gamma<$8\mev).
Also, the NA49 experiment reported evidence for an additional 
narrow exotic (B$=$1, S$=-$2) state, the $\Xi_5^{--}$, 
as well as the corresponding $\Xi_5^0$ state, with mass about 1862\mevcc~\cite{prl92:042003}.
More recently the H1 collaboration reported~\cite{hep-ex-0403017} 
a narrow ($\Gamma<30$\mev) exotic anti-charmed (B$=$1, C$=-$1) resonance, 
$\Thc$, with a mass of 3099$\pm$6\mevcc.
For these $\Theta^+$, $\Xi_5^{--}$ and $\Thc$ states the minimal quark content is ($udud\bar{s}$), ($dsds\bar{u}$) and
($udud\bar{c}$), respectively. 
These results have prompted a surge of pentaquark searches in experimental data
of many kinds, mostly with negative results~\cite{nullclaims}.
Several theoretical models~\cite{zp:a359:305,Marek,Jaffe} have been proposed to
describe possible pentaquark structure.
They predict that the lowest-mass states containing $u$, $d$ and $s$ quarks
should occupy a spin-1/2 anti-decuplet and octet.
The $\Theta_c^0$ pentaquark 
should be an isospin-zero member of the $\overline{6}$ representation of the 60-plet of SU(4)~\cite{Ma}.
 
The $\babar$ experiment~\cite{babardet} at the SLAC PEP-II $e^+e^-$ collider takes data  at 
center-of-mass energy of $\approx$10.58\gev. 
\babar\ is well suited to search for these states, since it provides excellent pion, 
kaon and proton identification and good tracking, with the result that excellent mass
resolution can be achieved. 
Charged particle tracks are measured by a five-layer silicon vertex tracker (SVT) and a 40-layer drift-chamber (DCH)
 located in a 1.5-T solenoidal magnetic field. 
Charged particles are identified by means of specific ionization ($dE/dx$) measurements in
the SVT and DCH, and from the pattern of Cherenkov photons in the
Cherenkov radiation detector (DIRC).

In addition to the inclusive searches for the strange and anti-charm pentaquark states
in $e^+e^-$ annihilations presented in this paper,
there is also an inclusive search for the $\Theta^+$ in electro- and hadro-production 
in the material of the \babar\ detector which shows no evidence for this state \cite{Jon}.  
Furthermore an exclusive search in $B^+$ decay to $p\bar{p}K^+$ final states 
for the isovector pentaquark candidate $\Theta^{*++}$ decaying into $pK^+$ in the mass range 
1.43 to 2.00$\,$GeV$/c^2$ sets limits on 
${\cal B}(B^+\rightarrow \Theta^{*++}\bar{p})\times{\cal B}(\Theta^{*++}\rightarrow pK^+)$  
at the $10^{-7}$ level \cite{tanya}. A review of these and other \babar\ hadronic results can be 
found in these proceedings \cite{dave}.

\section{Inclusive searches for strange and anti-charm pentaquark baryons} 
\label{sect:2} 

\begin{figure}
\includegraphics[height=.39\textheight]{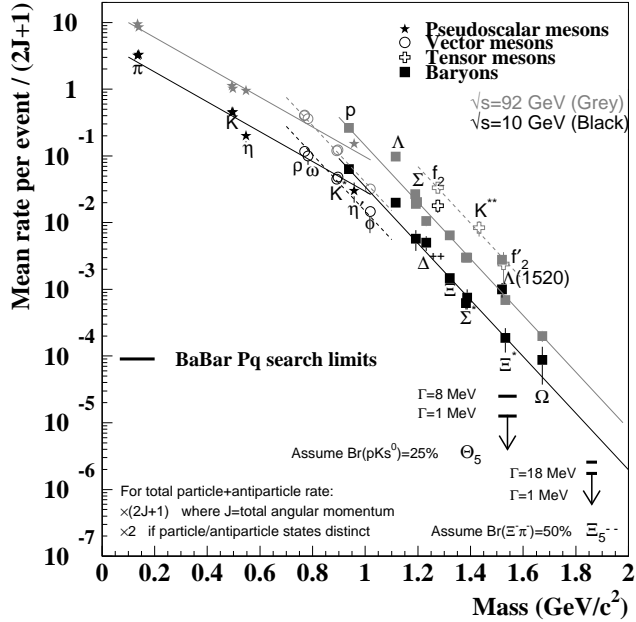} 
\caption{
Compilation of meson and baryon production rates in $e^+e^-$ 
annihilation~\cite{ref:pdg2002}
from experiments at the $Z^0$ (gray) and $\sqrt{s}\approx 10$\gev
(black) as a function of baryon mass.  
The vertical scale accounts for the number of spin and particle$+$antiparticle 
states, and the lines are chosen to guide the eye.  
The arrows indicate our upper limits on spin-1/2 $\Theta^+$ and
$\Xi_5^{--}$ pentaquark 
states, assuming the branching fractions shown, and are seen to lie below the 
solid line.
} 
\label{fig1}       
\end{figure}

Although experiments with a baryon in the beam or the target might seem to have some 
advantage in pentaquark production, $\epem$ interactions are also
known for democratic production of hadrons.
Mesons and baryons with non-zero charm and strangeness (up to three units) 
have been observed with production rates that appear to
depend on mass and spin, but not quark content (Fig.~\ref{fig1}).
If pentaquarks are produced similarly, then
one might expect a pentaquark rate as high as that for an ordinary
baryon of the same mass and spin.
The search for inclusive production of the pentaquark states \cite{charge}  
$\Theta^+$, $\Xi_5^{+}$, $\Xi_5^0$, $\Xi_5^{-}$, 
$\Xi_5^{--}$, $\Sigma_5^+$, $N_5^0$, $N_5^+$ and $\Thc$ 
has been performed with 123 fb$^{-1}$ of data recorded 
at or slightly below the $\Upsilon(4S)$ resonance~\cite{PRL95-042002-2005}. 
This paper will discuss only the $\Theta^+$, $\Xi_5^{--}$, $\Xi_5^0$ and 
$\Thc$ searches.

A search for the $\Theta^+$ is carried out in the decay mode $\Theta^+ \to p K^0_S$, where 
$K^0_S \rightarrow \pi^+ \pi^-$.
The expected $\Theta^+$ mass resolution is about $2$\mevcc. However 
no peak is seen at the expected mass but
a large signal at 2285 MeV/c$^2$ (with a mass resolution of  
6\mevcc at the $\Lambda_c$ mass) containing $\approx$98,000 entries from $\Lambda_c\to p K^0_S$ is observed.
This null result for a $\Theta^+$ mass  of 1540 MeV/c$^2$ is quantified by fitting the  
convolution of a Gaussian and a P-wave Breit-Wigner for the signal line-shape, and a seventh-order-polynomial 
times a threshold function for the background shape, 
to the $pK^0_S$ invariant-mass distribution in the interval from threshold to 1800\mevcc. 
Since the intrinsic width of the $\Theta^+$ has not been  
determined so far, width values of $\Gamma=$1\mev (for a narrow $\Theta^+$) and  
$\Gamma=$8\mev (best upper limit) are used, and the results quoted for each assumed width.  
The upper limit, at 95$\%$ confidence level, is determined for the number of produced  
pentaquarks per $e^+e^- \rightarrow hadrons$ event, and compared to the production  
rates of known  baryons, assuming  
${\cal B}(\Theta^+ \rightarrow pK^0_S)=25\%$. The measured upper limit values  
of 5$\times$10$^{-5}/$event ($\Gamma=$1\mev) and 11$\times$10$^{-5}/$event ($\Gamma=$8\mev) 
are between eight and 15 times lower than expected for conventional baryons, as shown in Fig.~\ref{fig1}. 

A search for the $\Xi_5^0$ and $\Xi_5^{--}$ resonances was performed using the decay chain
$\Xi_5\to \Xi^- \pi$, $\Xi^- \to \Lambda \pi^-$, $\Lambda \to p\pi^-$.
In each case, no peak is seen at the expected mass. In the $\Xi^-\pi^+$ spectrum, prominent
peaks for the $\Xi(1530)^0$ and $\Xi_c(2470)^0$ with $\approx$24,000 and $\approx$8,000 entries respectively, are seen. 
No structure is observed in the exotic $\Xi^{-}\pi^-$ spectrum. 
A linear function is used for the background, while the signal is modeled as described above.
The resolution function is derived from the $\Xi(1530)^0$ and $\Xi_c(2470)^0$ signals in
data and simulation, and is described by a Gaussian function with an RMS of 8\mevcc.
The fit is performed over a $\Xi^- \pi^-$ mass range from 1760 to 1960\mevcc.
As before, two different intrinsic widths of this pentaquark state are used, 
namely $\Gamma=$1\mev (narrow) and $\Gamma=$18\mev (best experimental upper limit) in order 
to determine 95$\%$ confidence level upper limit values for the production rate in $e^+e^-$ interactions.  
The values obtained, 0.74$\times$10$^{-5}/$event ($\Gamma=$1\mev) and 1.1$\times$10$^{-5}/$event ($\Gamma=$18\mev), are between 
four and six times lower than those for conventional baryons, as shown in Fig.~\ref{fig1},
assuming   ${\cal B}(\Xi_5^{--}\to \Xi^- \pi^-)=50\%$. 
It is not possible to determine the total production rate for the $\Xi_5^0$,
as its branching fraction to $\Xi^- \pi^+$ is unknown. 

\begin{figure}
\begin{center}
\begin{tabular}{ll}
\includegraphics[width=.49\textwidth]{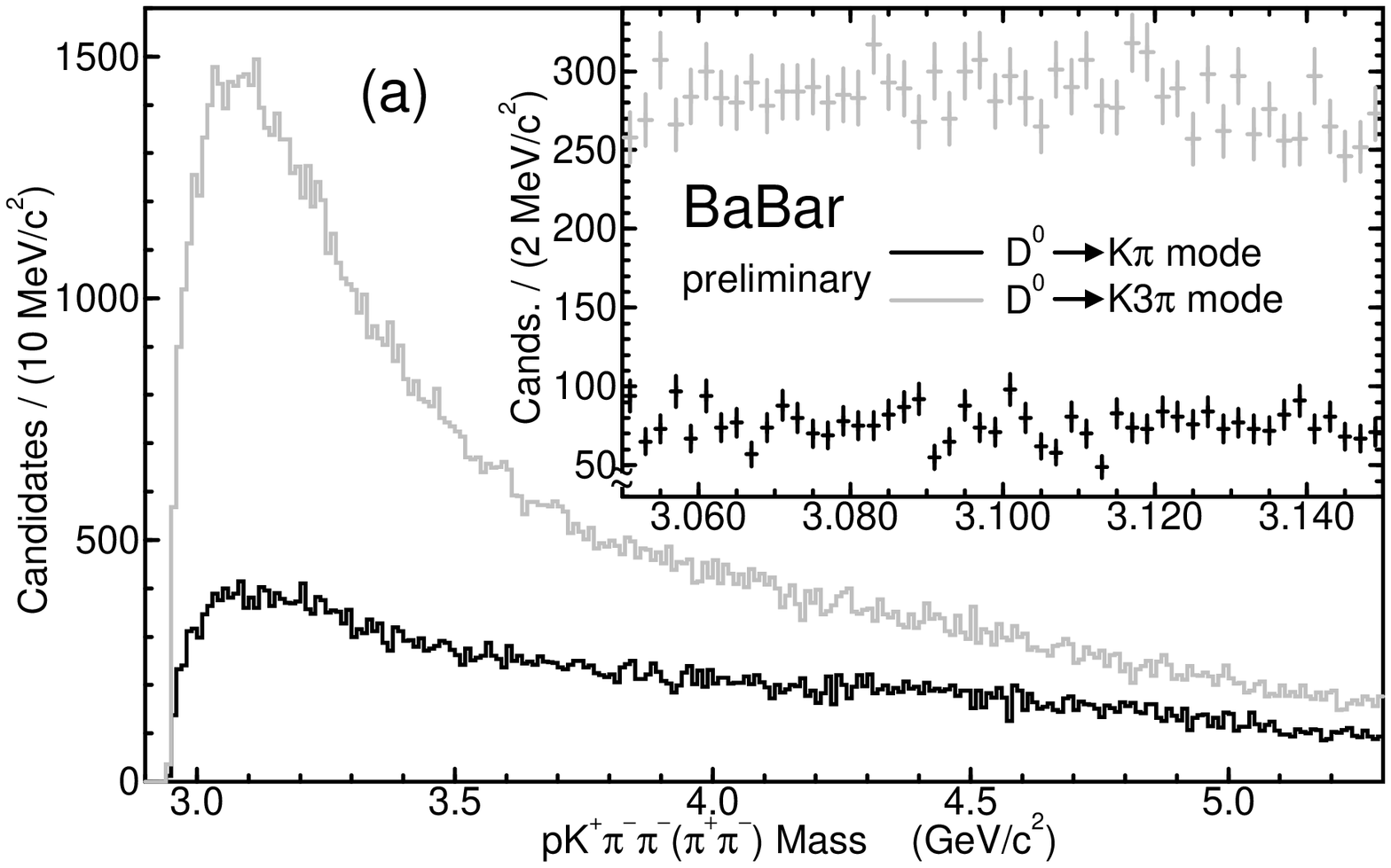} & \includegraphics[width=.49\textwidth,height=0.305\textwidth]{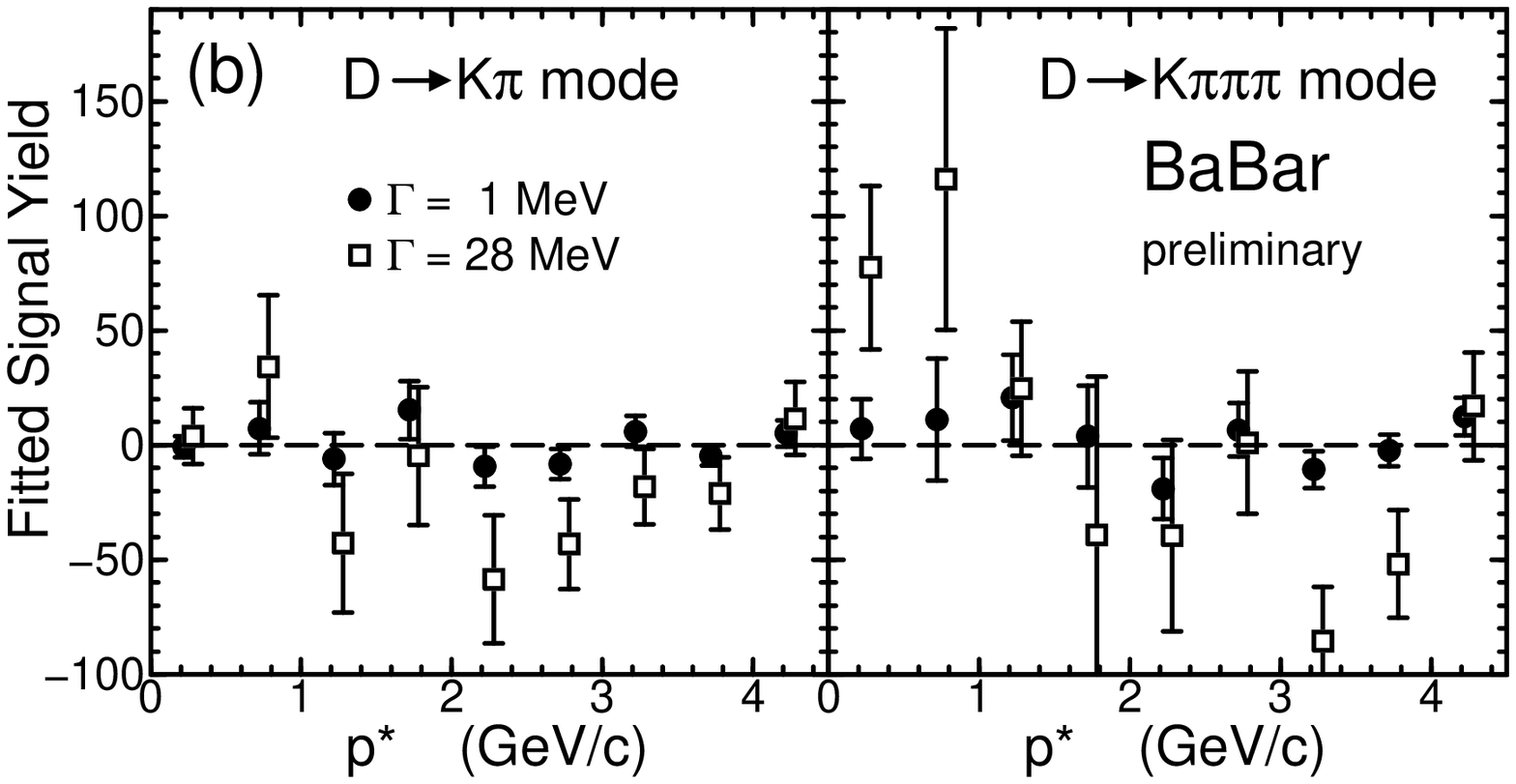} \\
\end{tabular}
\end{center}
\caption{(a) Invariant mass distributions of 
$D^{*-}p$ with $\bar{D}^0$ in the $K^+\pim$(black) and $K^+\pim\pim\pip$(gray) modes 
for combinations satisfying the criteria described in  
the text. The data are plotted for the full  
kinematically allowed $D^{*-}p$ range 
and, in the inset, with statistical uncertainties and a suppressed zero on the vertical axis,  
for the mass range in which the \Thc has been reported.
(b) The \Thc yields extracted from the fits to the (left) $pK^+\pim\pim$ and (right) $pK^+\pim\pim\pip\pim$ invariant mass
distributions, assuming a mass of 3099\mevcc and a natural
width of $\Gamma=1$\mev (circles) or $\Gamma=28$\mev (squares).} 
\label{fig2}       
\end{figure}   

We present preliminary results from a search for \Thc production 
performed using the decay mode, $\Thc \to p D^{*-}$,
where the $D^{*-}$ is reconstructed in the $\bar{D}^0 \pi^-$ decay mode,
and the $\bar{D}^0$ in the $K^+\pim$ and $K^+\pim\pim\pip$ modes;
the selection criteria are designed for high efficiency and
low bias against production mechanism. 
The selected candidates satisfy the following criteria: the $\bar{D}^0$ and $D^{*-}$ candidates 
both have mass values within 20\mevcc of the respective average peak values, and their
mass difference is within 2\mevcc of the average peak value.
These values are chosen to minimize the upper limit on the total cross
section after a momentum-dependent correction for efficiency.
A total of about 55,000 (73,000) true $D^{*-}\to K^+\pim(K^+\pim\pim\pip)\pi^-$
are present in the selected data over a background of 4,000 (62,000).

The invariant mass distributions for the \Thc candidates in the data are 
shown in Fig.~\ref{fig2}(a) for the two $\bar{D}^0$ decay modes separately.
The distributions show no narrow structure, and in particular they are
all quite smooth in the region near 3099\mevcc, as shown in the inset. 
To avoid sensitivity to the details of the production mechanism, 
the $p^*$ distribution is divided into nine intervals of width 500\mevc from 0 to 4.5\gevc,
and then a fit to the invariant mass distribution is carried out for each $p^*$ interval.

As for the previous searches a
P-wave Breit-Wigner line-shape convolved with the resolution function is used for the
signal modeling. The \Thc invariant mass resolution is obtained from simulation \cite{GEANT}, and is represented by
a sum of two Gaussian functions with a common center. The overall 
resolution, defined as the FWHM of the resolution function divided by 
2.355, averages 2.8~(3.0)\mevcc for the $K^+\pim$~($K^+\pim\pim\pip$) 
decay modes with a small dependence on $p^*$, the center-of-mass momentum of the \Thc.
The quoted results assume two widths, $\Gamma = 1$\mev,
corresponding to a very narrow state, and $\Gamma = 28$\mev,
corresponding to the width observed by H1.
The background is described in each $p^*$ bin by a threshold function.
Maximum likelihood fits are performed at several fixed \Thc mass values
in the vicinity of 3099\mevcc.
In every case the fit quality is good and the signal obtained is consistent with zero.
Results using different mass values are consistent within expected statistical
variations. Fixing the mass to the reported value of 3099\mevcc results in the
event yields shown in Fig.~\ref{fig2}(b).
There is no evidence of a pentaquark signal in any $p^*$ range,
and the roughly symmetric scatter
of the points about zero indicates low
momentum-dependent bias in the background function.

It is not possible to compare the sensitivity of our search with that of
H1, due to the presumably different production mechanism, and the fact
that there is no cross section measurement in Ref.~\cite{hep-ex-0403017}.
It is interesting to note that H1 selects about
3,500 $D^{*-}$ using only the $\bar{D}^0\to K^+\pim$ mode, with a background of about 1,500;
of these, about 550 appear in their $pD^{*-}$ mass plot, with mass
below 3.6\gevcc, resulting in a \Thc signal yield of 51$\pm$11 events.
Thus the observed \Thc account for roughly 1/70 of their $D^{*-}$
production.
In contrast, the \babar\ search reconstructs about 750,000 $D^{*-}$ above background, 128,000 of
which also have an identified proton in the event, but no
\Thc signal is observed.

It is also not possible to compare the production rate of \Thc to the rates measured for
ordinary charmed baryons, as only measurements for $\Lambda_c$(2285) and $\Sigma_c$(2455)
at much lower mass than \Thc are available at this point \cite{ref:pdg2002}. 
Furthermore, since ${\cal B}(\Thc \to pD^{*-})$, is not known 
it is not possible to set an upper limit on the total production rate of the \Thc.

\section{Conclusions} 
\label{sec:5} 
 
A large statistics high-resolution search for the reported  pentaquark states $\Theta^+$, $\Xi_5^{--}$, $\Xi_5^0$ 
and \Thc in $e^+e^-$ annihilations has been performed at $\babar$. Large signals for 
known baryon states have been found, but no excess is seen at the reported mass 
values for the pentaquark states. 
 
\section{Acknowledgments} 
The author is grateful for the extraordinary contributions of the PEP-II colleagues in achieving the excellent 
luminosity and machine conditions that have made this work possible. This work is supported by
DOE
and NSF (USA),
NSERC (Canada),
IHEP (China),
CEA and
CNRS-IN2P3
(France),
BMBF and DFG
(Germany),
INFN (Italy),
FOM (The Netherlands),
NFR (Norway),
MIST (Russia), and
PPARC (United Kingdom). 
Individuals have received support from CONACyT (Mexico), A.~P.~Sloan Foundation, 
Research Corporation,
and Alexander von Humboldt Foundation.



\begin{thebibliography}{}
\bibitem{charge} Charge-conjugate reactions are included implicitly throughout the paper.
\end{thebibliography}


\end{document}